\documentclass{article}
\usepackage{spconf,amsmath,graphicx}

\usepackage{amssymb}

\usepackage[T1]{fontenc}

\usepackage{psfrag}
\usepackage[table,xcdraw,dvipsnames]{xcolor}
\usepackage{subfig}

\usepackage{enumitem}

\usepackage{pgfplots}
\pgfplotsset{compat=newest}
\pgfplotsset{plot coordinates/math parser=false} 
\usepgfplotslibrary{groupplots}

\usepackage{tikz}
\usetikzlibrary{matrix}
\newlength\figureheight
\newlength\figurewidth 

\usepackage{multirow}

\usepackage[customcolors]{hf-tikz}

\tikzset{style gray/.style={
		set fill color=gray!50,
		set border color=white,
	},
	hor/.style={
		above left offset={-0.15,0.31},
		below right offset={0.15,-0.125},
		#1
	},
	ver/.style={
		above left offset={-0.1,0.3},
		below right offset={0.15,-0.15},
		#1
	}
}

\title{Improving Mesh-Based Motion Compensation by Using Edge Adaptive Graph-Based Compensated Wavelet Lifting for Medical Data Sets}
\name{Daniela Lanz and Andr\'{e} Kaup}
\address{Multimedia Communications and Signal Processing\\
	Friedrich-Alexander-Universit\"at Erlangen-N\"urnberg (FAU), Cauerstr. 7, 91058 Erlangen, Germany\\
	Email: \{Daniela.Lanz, Andre.Kaup\}@FAU.de\\}
\begin{document}
\ninept
\maketitle

\begin{abstract}
Medical applications like Computed Tomography (CT) or Magnetic Resonance Tomography (MRT) often require an efficient scalable representation of their huge output volumes in the further processing chain of medical routine. A downscaled version of such a signal can be obtained by using image and video coders based on wavelet transforms. The visual quality of the resulting lowpass band, which shall be used as a representative, can be improved by applying motion compensation methods during the transform. This paper presents a new approach of using the distorted edge lengths of a mesh-based compensated grid instead of the approximated intensity values of the underlying frame to perform a motion compensation. We will show that an edge adaptive graph-based compensation and its usage for compensated wavelet lifting improves the visual quality of the lowpass band by approximately 2.5~dB compared to the traditional mesh-based compensation, while the additional filesize required for coding the motion information doesn't change.
\end{abstract}
\begin{keywords}
Discrete Wavelet Transforms, Motion Compensation, Scalability, Signal Processing on Graphs, Computed Tomography
\end{keywords}
\section{Introduction}
\label{sec:intro}
Accessing, transmitting or storing medical data volumes can be a crucial task because the filesize often gets very large. Therefore downscaled versions of the original signal, e.g. in telemedical applications, can be very useful for tasks like browsing and fast previewing.

Subband coding provides an appropriate way to achieve scalability features without additional overhead \cite{lnt2011-23}. Using the wavelet transform, the signal is decomposed into a lowpass (LP) and a highpass (HP) band with the energy concentrated in the LP band. Blur and ghosting artifacts in the LP band caused by the motion of the CT or MRT volumes can be compensated by incorporating adequate motion compensation (MC) methods directly into the wavelet transform. This adaption to the signal leads to a higher visual quality of the LP band and a better energy compaction in fewer transform coefficients. While the first property is very important, when the LP band shall be used in medical applications as a downscaled representative, the second property results in a higher coding efficiency \cite{958672}.

In this paper we will present a novel approach to use common motion vector fields of a traditional mesh-based MC in a much more efficient way by exploiting the geometric structure of the underlying grid. This way we can guarantee that the number of bits required for encoding the motion information stays the same, while the visual quality of the LP band increases.

Section~\ref{sec:compWT} presents a brief overview of the compensated wavelet lifting, followed by a detailed description of the new edge adaptive graph-based approach for motion compensation in Section~\ref{sec:edge}. Simulation results are shown in Section~\ref{sec:results} followed by a short conclusion in Section~\ref{sec:conclusion}.

\section{Compensated Wavelet Lifting}
\label{sec:compWT}
\begin{figure}[tb]
	\begin{scriptsize}
		\centering
		\psfragscanon
		\psfrag{x}{$x$}
		\psfrag{y}{$y$}
		\psfrag{z}{$z$}
		\psfrag{t}{$t$}
		\psfrag{1}{$1$}
		\psfrag{2}{$2$}
		\psfrag{3}{$3$}
		\psfrag{4}{$4$}
		\psfrag{frac12}{$\frac{1}{2}$}
		\psfrag{2t-1}{$2t-1$}
		\psfrag{2t}{$2t$}
		\psfrag{f1}{$f_{1}$}
		\psfrag{f2}{$f_{2}$}
		\psfrag{f1}{$f_{1}$}
		\psfrag{f2}{$f_{2}$}
		\psfrag{f3}{$f_{3}$}
		\psfrag{f4}{$f_{4}$}
		\psfrag{f2t-1}{$f_{2t-1}$}
		\psfrag{f2t}{$f_{2t}$}
		\psfrag{HP1}{$\text{HP}_1$}
		\psfrag{LP1}{$\text{LP}_1$}
		\psfrag{HP2}{$\text{HP}_2$}
		\psfrag{LP2}{$\text{LP}_2$}
		\psfrag{HPt}{$\text{HP}_t$}
		\psfrag{LPt}{$\text{LP}_t$}
		\psfrag{MC}{MC}
		\psfrag{IMC}{$\text{MC}^{-1}$}
		\psfrag{prediction}{\textcolor{red}{prediction}}
		\psfrag{step}{\textcolor{red}{step}}
		\psfrag{update}{\textcolor{red}{update}}
		\psfragscanoff
		\includegraphics[width=\columnwidth]{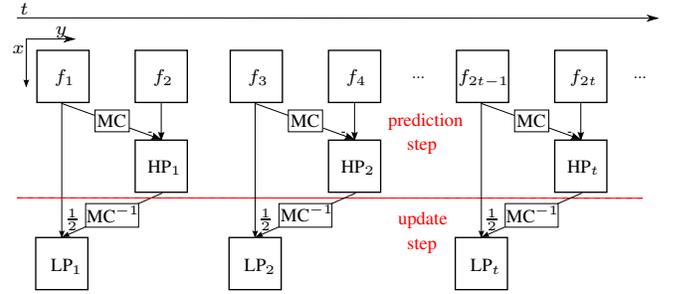}
		\caption{Compensated Haar lifting structure in temporal direction (MCTF).}
		\label{fig:MCTF}
	\end{scriptsize}
\end{figure}
By factorising the filter representation of the wavelet transform, it is possible to incorporate arbitrary compensation methods directly into the lifting structure~\cite{Sweldens1995}. Fig.~\ref{fig:MCTF} shows the lifting structure of the Haar wavelet and how it can be extended by a compensation method. The decomposition of the signal occurs in temporal direction, which is known as Motion Compensated Temporal Filtering (MCTF)~\cite{334985}. As Fig.~\ref{fig:MCTF} shows, the wavelet lifting consists of two steps, the prediction and the update step. The HP coefficients $\text{HP}_t$ are computed in the prediction step according to
\begin{equation}
\text{HP}_t = f_{2t}-\lfloor \mathcal{W}_{2t-1\rightarrow 2t}(f_{2t-1})\rfloor.
\end{equation}
Instead of a simple subtraction of the reference frame $f_{2t-1}$ from the current frame $f_{2t}$, a predictor, denoted by the warping operator $\mathcal{W}_{2t-1\rightarrow 2t}$, is used. This process is described by MC in Fig.~\ref{fig:MCTF}. To calculate the LP coefficients, the MC has to be inverted. This happens in the update step and is denoted by $\text{MC}^{-1}$. To achieve an equivalent wavelet transform, the index of $\mathcal{W}$ has to be reversed when calculating the LP coefficients
\begin{equation}
\text{LP}_t = f_{2t-1}+\lfloor \frac{1}{2}\mathcal{W}_{2t\rightarrow 2t-1}(\text{HP}_t)\rfloor.
\end{equation}
To avoid rounding errors, floor operators are applied in the transform \cite{647983}. Considering medical data, the reconstruction of the original signal without any loss is a very important aspect.

\section{Edge Adaptive Graph-Based Motion Compensation}
\label{sec:edge}

To reconstruct the original signal at the decoder side, it is necessary to encode the corresponding LP and HP bands as well as the motion information used for the MC and $\text{MC}^{-1}$. In traditional compensation methods, like block-based or mesh-based approaches \cite{lnt2012-40}, \cite{lnt2014-23}, the motion is stored in form of motion vector fields.

The novelty of this paper is to exploit the motion vector fields of a mesh-based motion estimation to get the displacements of a compensated grid and to use these displacements instead of the intensity values of the underlying frame for the motion compensation. Thereby the number of bits to code the motion information stays the same, while the visual quality of the LP band will increase by incorporating the geometric structure of the data.

\begin{figure}[tb]
	\centering
	\psfragscanon
	\psfrag{f1}{$f_{2t-1}$}
	\psfrag{f2}{$f_{2t}$}
	\psfrag{MVF}{motion vectors}
	\psfrag{upsample}{calculate positions}
	\psfrag{between}{between GPs}
	\psfrag{f1}{$f_{2t-1}$}
	\psfrag{f2}{$f_{2t}$}
	\psfrag{l2}{$e_b$}
	\psfrag{l3}{$e_i$}
	\psfrag{l4}{$\tilde{e}_i$}
	\psfrag{w}{$w(i)$}
	\psfrag{JP}{$\mathbf{J_P}$}
	\includegraphics[width=0.45\textwidth]{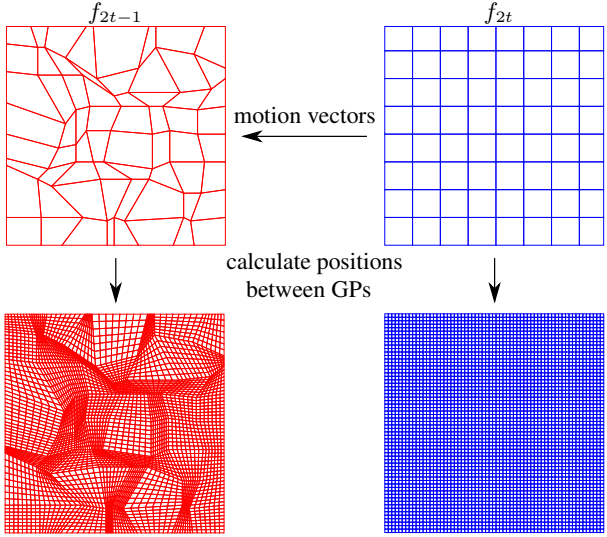}
	\psfragscanoff
	\caption{After the regular grid gets deformed by using the corresponding motion vectors, the subpixel positions have to be calculated. A traditional mesh-based MC uses the interpolated values of $f_{2t-1}$ at the subpixel positions, while the proposed method exploits the varying edge lengths of the compensated grid.}
	\label{fig:overview}
\end{figure}

Considering a 2-D mesh-based compensation which is calculated by putting a quadrilateral mesh of arbitrary grid size over frame $f_{2t-1}$ and deformed regarding frame $f_{2t}$, every motion vector of every single grid point (GP) is stored in the corresponding motion vector field. Then the missing positions of the pixels laying between the compensated GPs have to be calculated. An example for this process of deforming and upsampling a grid can be seen in Fig.~\ref{fig:overview}. Due to the deforming process, the links between the GPs are changing their length compared to the regular grid. 

\subsection{Graph-Based Motion Compensation}

\begin{figure}[tb]
	\centering
	\psfragscanon
	\psfrag{f1}{$f_{2t-1}$}
	\psfrag{f2}{$f_{2t}$}
	\psfrag{even}{even frame}
	\psfrag{odd}{odd frame}
	\includegraphics[width=0.37\textwidth]{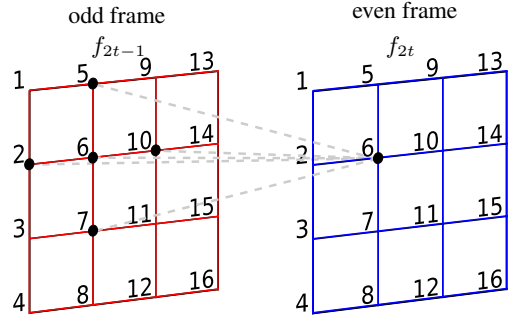}
	\psfragscanoff
	\caption{4-grid neighborhood for one single node of the even frame connected to the odd frame.}
	\label{fig:basic}
\end{figure}

A smart way to easily incorporate the varying edge lengths into the motion compensation is the graph-based wavelet lifting. As introduced in~\cite{narang2009lifting}, it is possible to perform a lifting-based wavelet transform on arbitrary graphs $G(\mathcal V,E)$, where $\mathcal{V}$ is the set of nodes, indexed as $1,2,3,...,N$ and $E$ is the set of links $e$ between the nodes. Every link is defined by a triplet $(i,j,w_{ij})$, where $i$ and $j$ are the start and end nodes respectively and $w_{ij}$ is the weight which has a value $\neq0$ if $i$ and $j$ are linked to each other. Also every node has a value which is listed in the vector $X$. For the graph-based wavelet transform a splitting of the nodes into even and odd subsets is required. As a consequence the corresponding adjacency matrix $\mathbf{A}$ has to be rearranged accordingly
\begin{equation}
X = \begin{pmatrix}
X_\text{even} \\ 
X_\text{odd}
\end{pmatrix} \quad 
\mathbf{A} = \begin{pmatrix}
\mathbf{F} & \mathbf{J} \\ \mathbf{K} & \mathbf{L}
\end{pmatrix},
\end{equation}
where the submatrices $\mathbf{F}$ and $\mathbf{L}$ contain edges, which connect nodes of same parity and the submatrices $\mathbf{J}$ and $\mathbf{K}$ contain all edges, which connect nodes of different parity. By applying 
\begin{equation}
\begin{aligned}
H &= X_\text{even}-\mathbf{J_P}\times X_\text{odd}\\
L &= X_\text{odd}+\mathbf{K_U}\times H
\end{aligned}
\label{eq_graph_trafo}
\end{equation}
we get the vector $H$, which contains the HP coefficients whereas the vector $L$ contains the LP coefficients. The matrices $\mathbf{J_P}$ and $\mathbf{K_U}$ are computed from $\mathbf{J}$ and $\mathbf{K}$ by assigning prediction and update weights depending on the desired application. Since the matrices $\mathbf{F}$ and $\mathbf{L}$ are not used in~(\ref{eq_graph_trafo}), a perfect splitting of nodes should be intended~\cite{hidane2013lifting}.

Considering images as graph signals as introduced in \cite{Lanz2016} and applying the graph-based wavelet transform, every single pixel of a frame has to be interpreted as a node. Accordingly the intensity values of the pixels are stored in vector $X$. To fulfill the constraint of perfect splitting, every frame gets assigned as even and odd according to its number of appearance in the sequence as shown in Fig.~\ref{fig:basic}. Then every node of an even frame gets linked by a previously defined neighborhood in the odd frame for constructing matrix $\mathbf{J}$ and vice versa for matrix $\mathbf{K}$. In Fig.~\ref{fig:basic} a 4-grid neighborhood is chosen which is shown for node 6 of the even frame connected to the corresponding nodes 2,5,6,7 and 10 in the odd frame. 
After a proper weighting of the referenced nodes, which is often measured as the spatial or photometric similarity between $i$ and $j$ \cite{6694319}, the degree matrices $\mathbf{D_{J/K}}$ of the weighted matrices $\mathbf{J}$ and $\mathbf{K}$ are computed. Using random walks on Graph $G$ delivers the Markov chain with the transition matrices
\begin{equation}
\begin{aligned}
\mathbf{P_J} & = \mathbf{D_J^{-1}J}\\
\mathbf{P_K} & = \mathbf{D_K^{-1}K}.
\end{aligned}
\end{equation}
According to \cite{lee2011multiscale} one element $p_{ij}$ of such a transition matrix equates the probability of being at node $j$ starting from node $i$. Therefore the transition matrices $\mathbf{P_J}$ and $\mathbf{P_K}$ are used as prediction matrix $\mathbf{J_P}$ and update matrix $\mathbf{K_U}$ respectively. Then a graph-based Haar wavelet transform can be carried out on images:
\begin{equation}
\begin{aligned}
H &= X_\text{even}-\lfloor\mathbf{J_P}\times X_\text{odd}\rfloor\\
L &= X_\text{odd}+\lfloor\frac{1}{2}\mathbf{K_U}\times H\rfloor.
\end{aligned}
\end{equation}
By rearranging the vectors $H$ and $L$ containing the transform coefficients to the original frame size, the HP and the LP bands are achieved again.

\subsection{Conversion of Motion Vector Fields into Adjacency Matrices}
\label{subsec:conversion}

\begin{figure}[tb]
	\centering
	\psfragscanon
	\psfrag{f1}{$f_{2t-1}$}
	\psfrag{f2}{$f_{2t}$}
	\psfrag{l2}{$e_b$}
	\psfrag{l3}{$e_i$}
	\psfrag{l4}{$\tilde{e}_i$}
	\psfrag{w(i)}{$w_{ij}(e)$}
	\psfrag{Jp}{$\mathbf{J_P}$}
	\includegraphics[width=0.5\textwidth]{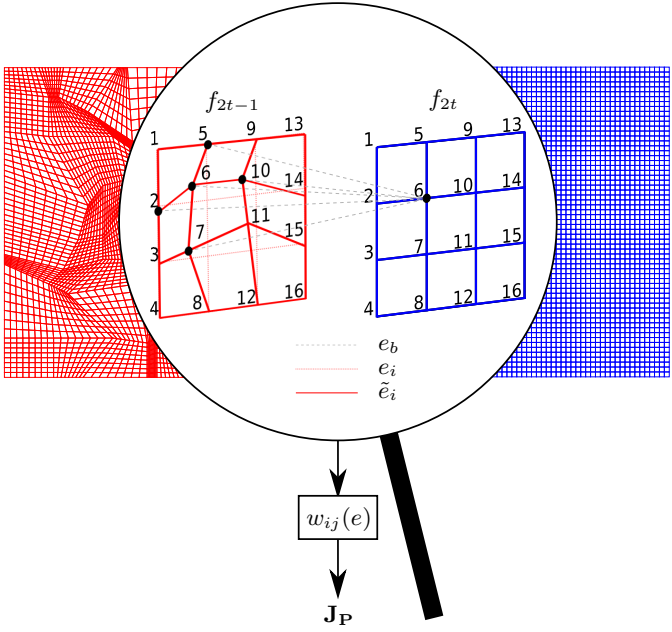}
	\psfragscanoff
	\caption{A zoom into the deformed grid of $f_{2t-1}$ and the regular grid  of $f_{2t}$ shows the various edges which are used to calculate the prediction matrix $\mathbf{J_P}$ according to the weighting function $w_{ij}(e)$ given in~(\ref{equ:w}).}
	\label{fig:zoom}
\end{figure}

The varying distances resulting from the process of deforming and upsampling the considered grid are used as edge weights in the prediction matrix $\mathbf{J_P}$ and the update matrix $\mathbf{K_U}$. Since the displacements in the signal can mainly be characterized as contraction and expansion of different kinds of tissue a proper weighting function is required which assigns higher values to decreased edge lengths and lower values to increased edge lengths.

Due to the deformed mesh and the graph connections there exists a plenty of varying distances which can be used for a proper weighting. As shown in Fig.~\ref{fig:zoom} mainly two different kinds of connections can be distinguished, namely:
\begin{itemize}[topsep=0mm]
	\itemsep0pt
	\item[-] inter-frame edges $e_b$: edges between the even and the odd frame
	\item[-] intra-frame edges $e_i$: edges inside the odd frame
\end{itemize}
Considering the change of their length if the underlying grid gets deformed a further differentiation can be introduced:
\begin{itemize}[topsep=0mm]
	\itemsep0pt
	\item[-] regular edges $e_i$: intra-frame edges on a regular grid
	\item[-] compensated edges $\tilde{e}_i$: intra-frame edges on a compensated grid
\end{itemize}
Hence, if $\tilde{e}_i$ is smaller than the corresponding $e_i$, a contraction of the specific tissue occurs. Otherwise the classification of the underlying movement is not unique, because if $\tilde{e}_i > e_i$, this can correspond to an expansion of the tissue or another kind of tissue could be referenced.
\pagebreak

Keeping these definitions in mind a weighting function $w_{ij}(e)$ can be formed which assigns higher weights if a contraction is identified:
\begin{equation}
\begin{aligned}
w_{ij}(e) &= \exp(-\frac{1}{2}\cdot (e_b^2+e^2))\cdot \exp(|e_b - \tilde{e}_i|),\\
e &= 
\begin{cases}
\tilde{e}_i & \text{if }  \tilde{e}_i<e_i\\
e_b & \text{else}. \\
\end{cases}
\end{aligned}
\label{equ:w}
\end{equation}
Other weighting functions would also be possible. After connecting every node of frame $f_{2t}$ by a previously defined neighborhood to the deformed frame $f_{2t-1}$ and using the above described weighting function, the resulting weights $w_{ij}$ are used for calculating matrix $\mathbf{J_P}$, as exemplarily shown in Fig.~\ref{fig:zoom} for node 6 of the even frame $f_{2t}$. Matrix $\mathbf{K_U}$ can easily be found by taking just the transpose of $\mathbf{J_P}$. 

A further property of this method is based on the fact that the upsampling process results in subpixel coordinates. A large number of investigations showed, that it is of advantage to round them to full pixel positions. Therefore it can occur that some end nodes are multiply referenced. This procedure contributes to a sharper differentiation of the classification of the underlying movement and therefore to a higher weighting of nodes belonging to a contraction.

\section{Simulation Results}
\label{sec:results}

\begin{table*}[tb]
	\centering
	\begin{tabular}{c|cccc|cccc}
		\multirow{2}{*}{} & \multicolumn{4}{c|}{PSNR LP {[}dB{]}} & \multicolumn{4}{c}{Mean energy HP} \\
		& \textit{thorax1} & \textit{thorax2} & \textit{thorax3} & \textit{head} & \textit{thorax1} & \textit{thorax2} & \textit{thorax3} & \textit{head} \\ \hline
		no MC & 43.85 & 42.86 & 43.37 & 33.74 & 2862.35 & 6103.93 & 3350.72 & 32101.65  \\
		block-based & 47.51 & 46.09 & 46.99 & 38.31 & 827.35 & 1794.19 & 908.34 & 6504.12  \\
		mesh-based & 49.92 & 47.90 & 49.40 & 40.31 & 767.66 & 2540.14 & 892.06 & 8256.79  \\
		edge adaptive graph-based& 51.62 & 50.16 & 51.53 & 44.17 & 767.24 & 2327.04 & 862.48 & 7635.86  \\ \hline
		$\Delta$: proposed to mesh-based & +1.70 & +2.26 & +2.13 & +3.86 & -0.42 & -213.10 & -29.58 & -620.93 
	\end{tabular}
	\caption{The table lists results regarding the visual quality and the mean energy for the considered sequences and various compensation methods. The values are averaged over the whole sequences, while the last row contains a delta between the proposed edge adaptive graph-based and the mesh-based approach.}
	\label{tab:results}
\end{table*} 

To evaluate the proposed edge adaptive graph-based wavelet transform three \textit{thorax} data sets and one \textit{head} CT data set were used. The \textit{thorax} sequences have a resolution of $512\times 512$~pixels at 12~bit per sample and describe a beating heart over time. \textit{thorax1} and \textit{thorax2} each consist of $10$ timesteps whereas \textit{thorax3} has $127$ frames in spatial direction. The \textit{head} sequence consists of $36$ frames in spatial direction at the same bit depth with a resolution of $448\times 448$ pixels.

For the simulation one Haar wavelet decomposition step is performed. Beside the proposed edge adaptive graph-based method the decomposition is done with a block-based, a mesh-based, and without a MC method. For the edge adaptive graph-based MC a 25-nearest-neighbor graph is used for connecting the particular even and odd frames. The MC of the corresponding grid is calculated according to~\cite{lnt2012-29} with a quadrilateral mesh and a grid size of $8\times 8$~pixels. For comparison the mesh-based approach is calculated with the same parameters, while the block-based approach uses a block size equal to the grid size and a search range of 8~pixels. To overcome the unconnected pixels appearing at the inversion in the update step of the block-based MC a nearest-neighbor interpolation was used. 

The averaged values regarding the visual quality and the mean energy for the considered sequences and various compensation methods can be seen in Table~\ref{tab:results}. As expected, the PSNR for all methods is significantly higher than a wavelet transform without any MC. As a consequence the mean energy of the corresponding HP band which can be regarded as the prediction error for a compensated wavelet transform decreases. By the last of row Table~\ref{tab:results} a $\Delta$ between the edge adaptive graph-based and the mesh-based approach is provided. The table proves that the edge adaptive graph-based MC outperforms the block-based and mesh-based approaches in terms of visual quality of the LP band and also the mesh-based approach regarding the mean energy of the HP band. 

\begin{table}[tb]
	\centering
	\begin{tabular}{c|cccc}
	\multirow{2}{*}{filesizes {[}kB{]}} & \textit{thorax1} & \textit{thorax2} & \textit{thorax3} & \textit{head} \\
	& \multicolumn{4}{c}{no MC}                                              \\ \hline
		LP                                  & 715.63           & 828.20           & 9235.87          & 2570.01        \\
		HP                                  & 697.42           & 946.60           & 8574.07          & 3002.13        \\
		MVF                                 & -                & -                & -                & -             \\ \hline
		$\Sigma$         					& 1413.05          & 1774.80          &  17809.94        & 5572.14      \\ 
		& 
		\multicolumn{4}{c}{}\\
		& \multicolumn{4}{c}{block-based}                                      										\\ \hline
		LP                                  & 859.64           & 985.98           & 10898.34         & 2849.85        \\
		HP                                  & 813.61           & 979.16           & 10356.63         & 2903.70        \\
		MVF                                 & 22.28            & 27.63            & 282.52           & 80.1152         \\ \hline
		$\Sigma$                            & 1695.52          & 1965.77          & 21537.54         & 5833.66\\ 
		& 
				\multicolumn{4}{c}{}\\
		& \multicolumn{4}{c}{mesh-based}                                       										\\ \hline
		LP                                  & 758.75           & 883.86           & 9633.65          & 2759.30        \\
		HP                                  & 733.48           & 916.73           & 9337.15          & 2841.11        \\
		MVF                                 & 20.18            & 24.34            & 254.20           & 62.66          \\ \hline
		$\Sigma$                            & 1512.41          & 1824.93          & 19225.00         & 5663.06 \\ 
		& 
				\multicolumn{4}{c}{}\\
		& \multicolumn{4}{c}{edge adaptive graph-based}                                      				   		\\ \hline
		LP                                  & 857.40           & 950.61           & 10782.08         & 2825.85        \\
		HP                                  & 767.63           & 905.40           & 9555.49          & 2773.93        \\
		MVF                                 & 20.18            & 24.34            & 254.20           & 62.66          \\ \hline
		$\Sigma$                            & 1645.21          & 1880.35          & 20591.77         & 5662.44       
	\end{tabular}
	\caption{The table summerizes the overall filesizes of the single subband volumes and the required motion vector fields (MVF) for the considered sequences and various compensation methods.}
	\label{tab:filesizes}
\end{table}

To evaluate the compressibility, the resulting subbands are coded losslessly using the wavelet-based volume coder JPEG~2000. In this simulation the OpenJPEG \cite{openjpeg} implementation was used. The motion vector fields were coded using the QccPack library \cite{fowler2000qccpack}. For both subband volumes 4 further wavelet decomposition steps in $xy$-direction were applied. Table~\ref{tab:filesizes} lists the filesizes in kilobytes from lossless coding of the LP, HP and the corresponding motion vector fields and the sum of them for the considered sequences and various compensation methods. According to \cite{lnt2014-23} a wavelet transform without any MC is recommended, when the quality of the LP band is not of interest. This is confirmed by the first part of Table~\ref{tab:filesizes}, where the overall filesize for a wavelet transform without any MC is much lower compared to the other approaches. The reasons for this behavior are the correlated noisy structures that can be exploited by the traditional wavelet transform without a MC. However, if a compensated wavelet transform is applied, it is not possible to exploit the structures of the noise anymore. Therefore the particular filesizes of the single subbands increase. And in addition the corresponding motion vector fields have also to be coded and contribute to the overall filesizes. But when the LP band is used as a scalable representation, the quality is important which can be increased by various compensation methods. The mesh-based method achieves a higher PSNR and a smaller filesize compared to the block-based method. The edge adaptive graph-based approach achieves a further improvement of the visual quality by 2.5~dB on average, while the overall filesize decreases compared to the block-based MC but increases slightly compared to the mesh-based MC. However, due to the fact that the edge adaptive graph-based approach uses the same motion vector fields as the mesh-based approach, the bits needed to code the motion vector fields are exactly the same. This can be seen in Table~\ref{tab:filesizes} by the rows regarding the filesizes of the motion vector fields. 

A low mean energy of the HP band indicates a good MC of the odd frame. For a high quality LP band not only a good MC but also a proper $\text{MC}^{-1}$ is required. The block-based approach contains annoying block artifacts in the LP band because of unconnected pixels. There exist various ways to conceal this erroneous structures like interpolation methods on the motion vector fields as proposed in \cite{bozinovic2005} or extrapolation methods like FSE as shown in \cite{lnt2013-18}. In contrast the mesh-based approach has for the sequences \textit{thorax2} and \textit{head} a higher mean energy of the HP band, but the PSNR of the LP band is better compared to the block-based approach according to Tabel~\ref{tab:results}. The $\text{MC}^{-1}$ of the mesh-based approach accepts an error by using only an approximation term instead of the quite complex inversion in the update step but works even better than the block-based approach. However, the proposed edge adaptive graph-based method results in a quite low mean energy of the HP band and ends up in a high quality LP band at the same time. The inversion of the edge adaptive graph-based MC works very simple by just taking the transpose of the prediction matrix. 

\section{Conclusion}
\label{sec:conclusion}
In this paper a novel edge adaptive graph-based compensated wavelet transform for medical data sets was introduced. To avoid the usage of interpolated intensity values for motion compensation, a new approach of exploiting common mesh-based motion vector fields is proposed. By incorporating the geometric structure of the data regarding the varying edge lengths of the compensated grid, a high quality LP band and a HP band with a low mean energy can be achieved. Since the motion information used for compensation is exactly the same like the mesh-based approach uses, the number of bits needed to code the motion vectors fields stays the same. As the overall filesize is slightly larger compared to the mesh-based approach, further work aims at the investigation of a proper coding of edge adaptive graph-based compensated subbands. Also the suitability of different weighting functions should be examined.

\section*{ACKNOWLEDGEMENT}
We gratefully acknowledge that this work has been supported by the Deutsche Forschungsgemeinschaft (DFG) under contract number KA 926/4-3.


\bibliographystyle{IEEE}
\bibliography{Literatur}

\begin{thebibliography}{10}

\bibitem{lnt2011-23}
J.~Garbas, B.~Pesquet-Popescu, and A.~Kaup,
\newblock ``Methods and tools for wavelet-based scalable multiview video
  coding,''
\newblock {\em IEEE Transactions on Circuits and Systems for Video Technology},
  vol. 21, no. 2, pp. 113--126, February 2011.

\bibitem{958672}
A.~Secker and D.~Taubman,
\newblock ``Motion-compensated highly scalable video compression using an
  adaptive 3d wavelet transform based on lifting,''
\newblock in {\em Proc. IEEE Int. Conf. on Image Processing (ICIP)},
  Thessaloniki, Greece, Oct 2001, vol.~2, pp. 1029--1032 vol.2.

\bibitem{Sweldens1995}
W.~Sweldens,
\newblock ``Lifting scheme: a new philosophy in biorthogonal wavelet
  constructions,''
\newblock in {\em Proc. SPIE International Symposium on Optical Science,
  Engineering, and Instrumentation}, 1995, vol. 2569, pp. 68--79.

\bibitem{334985}
J.~R. Ohm,
\newblock ``Three-dimensional subband coding with motion compensation,''
\newblock {\em IEEE Transactions on Image Processing}, vol. 3, no. 5, pp.
  559--571, Sep 1994.

\bibitem{647983}
A.R. Calderbank, I.~Daubechies, W.~Sweldens, and B.-L. Yeo,
\newblock ``Lossless image compression using integer to integer wavelet
  transforms,''
\newblock in {\em Proc. IEEE Int. Conf. on Image Processing (ICIP)}, Oct 1997,
  vol.~1, pp. 596--599.

\bibitem{lnt2012-40}
W.~Schnurrer, J.~Seiler, E.~Wige, and A.~Kaup,
\newblock ``Analysis of displacement compensation methods for wavelet lifting
  of medical 3-d thorax ct volume data,''
\newblock in {\em Proc. IEEE Int. Conf. on Visual Communication and Image
  Processing (VCIP)}, San Diego, CA, USA, November 2012, pp. 1--6.

\bibitem{lnt2014-23}
W.~Schnurrer, T.~Richter, J.~Seiler, C.~Herglotz, and A.~Kaup,
\newblock ``3-d mesh compensated wavelet lifting for 3-d+t medical ct data,''
\newblock in {\em Proc. IEEE Int. Conf. on Image Processing (ICIP)}, Paris,
  France, October 2014, pp. 3631 -- 3635.

\bibitem{narang2009lifting}
S.~K. Narang and A.~Ortega,
\newblock ``Lifting based wavelet transforms on graphs,''
\newblock in {\em Proc. IEEE Asia-Pacific Signal and Information Processing
  Association, Annual Summit and Conference (APSIPA ASC)}, Sapporo, Japan,
  October 2009, pp. 441--444.

\bibitem{hidane2013lifting}
M.~Hidane, O.~L{\'e}zoray, and A.~Elmoataz,
\newblock ``Lifting scheme on graphs with application to image
  representation,''
\newblock in {\em Proc. IEEE Glob. Conf. on Signal and Information Processing
  (GlobalSIP)}, Austin, Texas, USA, 2013, pp. 431--434.

\bibitem{Lanz2016}
D.~Lanz and A.~Kaup,
\newblock ``Grap-based compensated wavelet lifting for 3-d medical ct data,''
\newblock to appear in Proc. IEEE Picture Coding Symposium (PCS), Nuremberg,
  Germany, December 2016.

\bibitem{6694319}
Y.-H. Narang, S.K.~Chao and A.~Ortega,
\newblock ``Critically sampled graph-based wavelet transforms for image
  coding,''
\newblock in {\em Proc. IEEE Asia-Pacific Signal and Information Processing
  Association (APSIPA)}, Kaohsiung, Taiwan, Oct 2013, pp. 1--4.

\bibitem{lee2011multiscale}
J.~D. Lee and M.~Maggioni,
\newblock ``Multiscale analysis of time series of graphs,''
\newblock in {\em Proc. Int. Conf. on Sampling Theory and Applications
  (SampTA)}, Singapore, May 2011.

\bibitem{lnt2012-29}
W.~Schnurrer, T.~Richter, J.~Seiler, and A.~Kaup,
\newblock ``Analysis of mesh-based motion compensation in wavelet lifting of
  dynamical 3-d+t ct data,''
\newblock in {\em Proc. IEEE Int. Workshop on Multimedia Signal Processing
  (MMSP)}, Banff, Canada, September 2012, pp. 152--157.

\bibitem{openjpeg}
A.~Descampe, F.~Devaux, H.~Drolon, D.~Janssens, and Y.~Verschueren,
\newblock ``Openjpeg~2.0.0,'' http://www.openjpeg.org, Nov. 2012.

\bibitem{fowler2000qccpack}
J.E. Fowler,
\newblock ``Qccpack: An open-source software library for quantization,
  compression, and coding,''
\newblock in {\em Proceedings Applications of Digital Image Processing XXIII},
  San Diego, CA, USA, Aug. 2000, vol. 4115, pp. 294--301.

\bibitem{bozinovic2005}
N.~Bozinovic, J.~Konrad, W.~Zhao, and C.~Vazquez,
\newblock ``On the importance of motion invertibility in mctf/dwt video
  coding,''
\newblock Philadelphia, PA, USA, Mar. 2005, pp. 49--52.

\bibitem{lnt2013-18}
W.~Schnurrer, J.~Seiler, and A.~Kaup,
\newblock ``Improving block-based compensated wavelet lifting by reconstructing
  unconnected pixels,''
\newblock in {\em Proc. Int. Symp. on Signals, Circuits and Systems (ISCAS)},
  Iasi, Romania, July 2013, pp. 1--4.

\end{thebibliography}

\end{document}